\documentclass[aps,prev,twocolumn,superscriptaddress,preprintnumbers,floatfix,nofootinbib]{revtex4-1}
\pdfoutput=1
\usepackage{graphicx}
\usepackage{bm}
\usepackage{times}
\usepackage{slashed}
\usepackage{color}
\usepackage{aas_macros}
\usepackage{slashed}
\usepackage{lipsum}
\usepackage{subfigure}
\usepackage{multirow}
\usepackage{amsmath}
\usepackage{array} 
\usepackage{varwidth} 

\newcommand{\be}{\begin{equation}}
\newcommand{\ee}{\end{equation}}
\newcommand{\bea}{\begin{eqnarray}}
\newcommand{\eea}{\end{eqnarray}}

\begin{document}
\title{Pulsar TeV Halos Explain the TeV Excess Observed by Milagro}
\author{Tim Linden}
\email{linden.70@osu.edu}
\affiliation{Center for Cosmology and AstroParticle Physics (CCAPP), and \\ Department of Physics, The Ohio State University Columbus, OH, 43210 }
\author{Benjamin J. Buckman} 
\email{buckman.12@osu.edu}
\affiliation{Center for Cosmology and AstroParticle Physics (CCAPP), and \\ Department of Physics, The Ohio State University Columbus, OH, 43210 }

\begin{abstract}
Milagro observations have found bright, diffuse TeV emission concentrated along the galactic plane of the Milky Way. The intensity and spectrum of this emission is difficult to explain with current models where \mbox{$\gamma$-ray} production is dominated by hadronic mechanisms, and has been named the ``TeV excess". We show that TeV emission from pulsars naturally explains this excess. In particular, recent observations have detected ``TeV halos" surrounding pulsars that are either nearby or particularly luminous. Here, we show that the full population of Milky Way pulsars will produce diffuse TeV emission concentrated along the Milky Way plane. The total $\gamma$-ray flux from TeV halos is expected to exceed the hadronic  $\gamma$-ray flux at energies above $\sim$500~GeV. Moreover, the spectrum and intensity of TeV halo emission naturally matches the TeV excess. If this scenario is common to all galaxies, it will decrease the contribution of star-forming galaxies to the IceCube neutrino flux. Finally, we show that upcoming HAWC observations will resolve a significant fraction of the TeV excess into individual TeV halos, conclusively confirming, or ruling out, this model.

\end{abstract}

\maketitle


Observations by the Milagro telescope have detected bright, diffuse TeV $\gamma$-ray emission emanating from the Milky Way galactic plane~\citep{Atkins:2005wu}. Early analyses considered the region of interest (ROI) 40$^\circ$~$<$~$\ell$~$<$~100$^\circ$ and $|b|$~$<$~5$^\circ$, finding a diffuse $\gamma$-ray flux of (6.4 $\pm$ 1.4 $\pm$ 2.1) $\times$ 10$^{-11}$~cm$^{-2}$s$^{-1}$sr$^{-1}$ at a median energy of 3.5~TeV. This exceeds the diffuse $\gamma$-ray flux predicted from local cosmic-ray measurements by nearly an order of magnitude~\citep{Prodanovic:2006bq, Evoli:2007iy}, and has thus been dubbed the ``TeV excess"~\citep{Prodanovic:2006bq}. Subsequent observations detected this emission at the even higher energy of 15~TeV, subdividing the ROI into several sub-regions along the galactic plane in a smaller latitude range $|b|$~$<$~2$^\circ$~\citep{Abdo:2008if}. This excess was not observed at lower energy in subsequent measurements by the ARGO-YBJ collaboration~\citep{Bartoli:2015era}, a finding that is only consistent with the Milagro measurement if the spectrum of diffuse hadronic $\gamma$-ray emission at TeV energies becomes significantly harder than the $\alpha$~=~-2.7 spectrum found at GeV energies. 

Two classes of models have been posited to explain the Milagro emission. The first utilizes standard cosmic-ray production models, which are dominated by primary cosmic-ray protons accelerated by supernovae. Fitting these models to the TeV excess requires modifying cosmic-ray propagation. Work by the {\tt Galprop} and Milagro teams noted that, by relaxing constraints from local cosmic-ray measurements and instead normalizing the Milagro flux to the observed Energetic Gamma Ray Experiment Telescope (EGRET) excess~\citep{1997ApJ...481..205H}, the majority of the TeV excess could be explained by hadronic diffuse emission~\citep{Abdo:2008if}.  However, three subsequent observations have challenged this interpretation. First, the Fermi-LAT showed that the EGRET excess was an instrumental artifact~\citep{2009arXiv0907.0294P}, removing the impetus for renormalizing the diffuse $\gamma$-ray spectrum. Second, AMS-02 measurements of the local cosmic-ray proton spectrum have strongly constrained any hardening of the local cosmic-ray proton spectrum~\citep{Aguilar:2015ooa}. Third, ARGO-YBJ observations did not find a significant excess at lower energies, which would necessitate a sharp (and unphysical) break in the hadronic $\gamma$-ray spectrum~\citep{Bartoli:2015era}.

An alternative method of tuning cosmic-ray diffusion employed spatially variable cosmic-ray diffusion models to avoid constraints from the locally observed cosmic-ray density~\citep{Gaggero:2014xla}. In these models, the energy index of the diffusion coefficient increases with galactocentric radius. This hardens the spectrum of the hadronic $\gamma$-rays near the Galactic center, while keeping the local cosmic-ray spectrum consistent with observations. In addition to explaining the TeV excess~\citep{Gaggero:2015xza}, it is argued that this model provides a better fit to the low-energy diffuse $\gamma$-ray emission observed by the Fermi-LAT~\citep{Gaggero:2014xla}. However, standard models of cosmic-ray diffusion also explain the diffuse GeV $\gamma$-ray emission to within systematic errors~\citep{Ackermann:2012pya}. Thus, while this model can explain the Milagro data, it is best thought of as a fit to the TeV excess, rather than a result that is strongly motivated by external theory or observation. 

The second class of models produces additional TeV emission from a population of individually sub-threshold point sources, which may have a harder TeV $\gamma$-ray spectrum than the diffuse emission~\citep{Prodanovic:2006bq}. This new component would exceed the intensity of the soft-spectrum diffuse $\gamma$-ray flux at high energies, while remaining subdominant at GeV energies. However, up until this point, no source class has been uncovered that can reasonably produce both the spectrum and intensity of Milagro observations.  

We show that TeV $\gamma$-ray emission from young and middle aged pulsars \emph{must} produce TeV emission with the intensity and spectrum observed by Milagro. This work builds upon existing observations by Milagro, the High Altitude Water Cherenkov (HAWC) telescope~\citep{Abeysekara:2013tza}, and the High Energy Stereoscopic System (H.E.S.S.)~\citep{Aharonian:2004bw}. Each telescope has observed bright, spatially extended, emission coincident with energetic pulsars. We show that this emission is typical of all pulsars, and that the total emission from the population of individually sub-threshold pulsars produces a diffuse $\gamma$-ray emission matching the intensity of the TeV excess. Moreover, the very hard spectrum of TeV emission from pulsars makes Milagro observations compatible with constraints from ARGO-YBJ and the Fermi-LAT. We show that future observations by the HAWC telescope will resolve many of the sources responsible for the TeV excess, allowing us to confirm or rule out this model in the near future.

\emph {Observations of TeV Halos ---} Recent observations by Milagro~\citep{2009ApJ...700L.127A}, HAWC~\citep{Abeysekara:2017hyn}, and HESS~\citep{Abdalla:2017vci} have found a number of TeV $\gamma$-ray sources coincident with bright Australian Telescope National Facility (ATNF) pulsars~\citep{Manchester:2004bp}. These sources have several key features. First, they have a hard $\gamma$-ray spectrum ($\sim$E$^{-2.2}$) consistent with $\gamma$-ray emission generated by the inverse-Compton scattering of the same e$^+$e$^-$ population that produces the synchrotron emission observed in x-ray pulsar wind nebulae (hereafter, PWN)~\citep{Yuksel:2008rf, Hooper:2017gtd}. Second, they are bright, with a $\gamma$-ray intensity which indicates that $\sim$10\% of the pulsar spin-down power is converted into e$^+$e$^-$ pairs~\citep{Hooper:2017gtd}. Third, they are large, with a radial extent that increases as a function of the pulsar age and typically extends to $\gtrsim$10~pc for pulsars of ages $\gtrsim$100~kyr. This final observation is difficult to explain in the context of PWN, whose size can be accurately modeled by calculating an equilibrium distance between the energy density of the relativistic pulsar wind (which decreases as a function of the pulsar age), and the interstellar medium. Observed PWN have radial extents of $\sim$1~pc, filling a volume three orders of magnitude smaller than the TeV emission. The scale of the TeV emission requires a new physical model, and these sources have thus been termed ``TeV halos"~\citep{Linden:2017vvb}.


Because the rotational kinetic energy of the pulsar is the ultimate power-source of all PWN and TeV halo emission, the high luminosity of TeV halos place strong constraints on every phase of $\gamma$-ray generation. First, pulsars must efficiently convert a significant fraction of their total spin-down power into e$^+$e$^-$ pairs. Second, high-energy ($\gtrsim$10~TeV) e$^+$e$^-$ electrons and positrons must lose a significant fraction of their total energy to synchrotron or inverse-Compton cooling before exiting the TeV halo.  Third, inverse-Compton scattering must constitute a reasonable fraction of the total e$^+$e$^-$ energy loss rate. In the case of the well-studied Geminga pulsar, the best fit models indicate that between 7-29\% of the total pulsar spin-down energy is converted into e$^+$e$^-$ pairs, that e$^+$e$^-$ with energies $\gtrsim$10~TeV lose more than 85\% of their total energy before leaving the TeV halo, and that approximately half of that cooling proceeds via inverse-Compton scattering~\citep{Hooper:2017gtd}. 

At present, TeV halos are observed from only a handful of pulsars that are either highly energetic, or relatively nearby. However, preliminary evidence strongly suggests that TeV halos are a generic feature of all young and middle-aged pulsars. Momentarily constraining our analysis to pulsars older than 100~kyr, in order to remove any contamination from supernova remnants, we note that the ATNF catalog lists 57 pulsars with reliable distance estimates that overlap the HAWC field of view. Assuming that the TeV halo flux of each system scales linearly with the pulsar spin-down energy and is inversely proportional to the square of its distance, we can produce a ranked list of the expected TeV halo flux.  Five of the seven brightest systems are currently detected by HAWC, while none of the dimmer systems have an observed TeV association~\citep{Linden:2017vvb}. This is compatible with the expected flux of each system, calculated by normalizing the TeV halo efficiency to Geminga. If TeV halos are, in fact, a generic feature of young and middle-aged pulsars, the ensemble of these dimmer systems is expected to provide a bright TeV $\gamma$-ray flux which cannot (at present) be separated into individual TeV halos.

\vspace{0.2cm}
\emph {Models for the Hadronic Gamma Rays ---} To determine whether TeV halos can produce the diffuse $\gamma$-ray emission observed by Milagro, we use a background model for the diffuse $\gamma$-ray emission from standard astrophysical processes. Specifically, we utilize the ensemble of 128 models developed by the Fermi-LAT collaboration to explain the diffuse GeV $\gamma$-ray flux~\citep{Ackermann:2012pya}. These models employ the cosmic-ray propagation code {\tt Galprop}, which physically models the production, propagation, and diffuse emission of cosmic-rays throughout the Milky Way~\citep{Strong:1998fr, Strong:2009xj}. The primary cosmic-rays injected in these models are dominated by protons accelerated by supernovae, and thus we refer to this emission as a  ``hadronic background", even though it includes some emission from primary and secondary leptons. As opposed to the models developed by~\citep{Gaggero:2014xla, Gaggero:2015xza}, these models are not tuned to explain the TeV excess, making them a natural choice to investigate the TeV halo contribution to the Milagro data. Additionally, these models span a wide parameter space of reasonable diffusion parameters, diffusion halo heights, molecular gas temperature models, and supernova injection morphologies. However, it is worth noting that, because these models are tuned to fit the diffuse emission spectrum observed by the Fermi-LAT, and include no additional physical features at TeV energies, they produce fairly similar predictions for the diffuse $\gamma$-ray flux. 

While the output from all 128 {\tt Galprop} models is publicly available, these simulations were terminated at a $\gamma$-ray energy of 1~TeV. We re-run each {\tt Galprop} model, extending the maximum cosmic-ray proton energy to 10~PeV, and the maximum $\gamma$-ray energy to 100~TeV. The maximum energy in our model does not affect our results. As there is no evidence (outside of the TeV excess) for new cosmic-ray physics at TeV energies, this provides a straightforward extrapolation of our understanding of GeV cosmic-ray physics to TeV energies. For each {\tt Galprop} model, we calculate the average hadronic $\gamma$-ray flux in both Milagro ROIs, finding that they significantly underproduce the observed signal. 

\vspace{0.2cm}
\emph {Models for the TeV Halo Flux ---} Because {\tt Galprop} produces a physical model for the diffuse $\gamma$-ray emission from cosmic-ray propagation, {\tt Galprop} self-consistently calculates the Milky Way supernova rate in the Milagro ROI. This provides a normalization for the pulsar birth rate, which we use to calculate the TeV halo formation rate. We assume that all supernovae produce pulsars. This is a mild overestimate -- but it affects the TeV halo intensity linearly and is degenerate with several assumptions in this study. While pulsars obtain a significant natal kick due to asymmetries in the supernova explosion~\citep{Hobbs:2005yx}, a typical kick of $\sim$400~km/s moves a pulsar only $\sim$40~pc over the 100~kyr period where a pulsar produces most of its TeV halo emission. We ignore this effect.

The injected cosmic-ray proton luminosity in our {\tt Galprop} models lies between \mbox{0.69 - 1.2~$\times$~10$^{40}$~erg~s$^{-1}$}. Assuming that each supernova injects 10$^{51}$~erg in kinetic energy and 10\% of this energy is emitted in cosmic-ray protons, this corresponds to a supernova rate in the Milagro ROI of 0.0021---0.0037~yr$^{-1}$, and a total Milky Way rate of $\sim$0.015~yr$^{-1}$. This matches observations that find the Milky Way supernova rate to be 0.019$\pm$0.011~yr$^{-1}$~\citep{Diehl:2006cf}. 
\\
\begin{figure}[tbp]
\centering
\includegraphics[width=.48\textwidth]{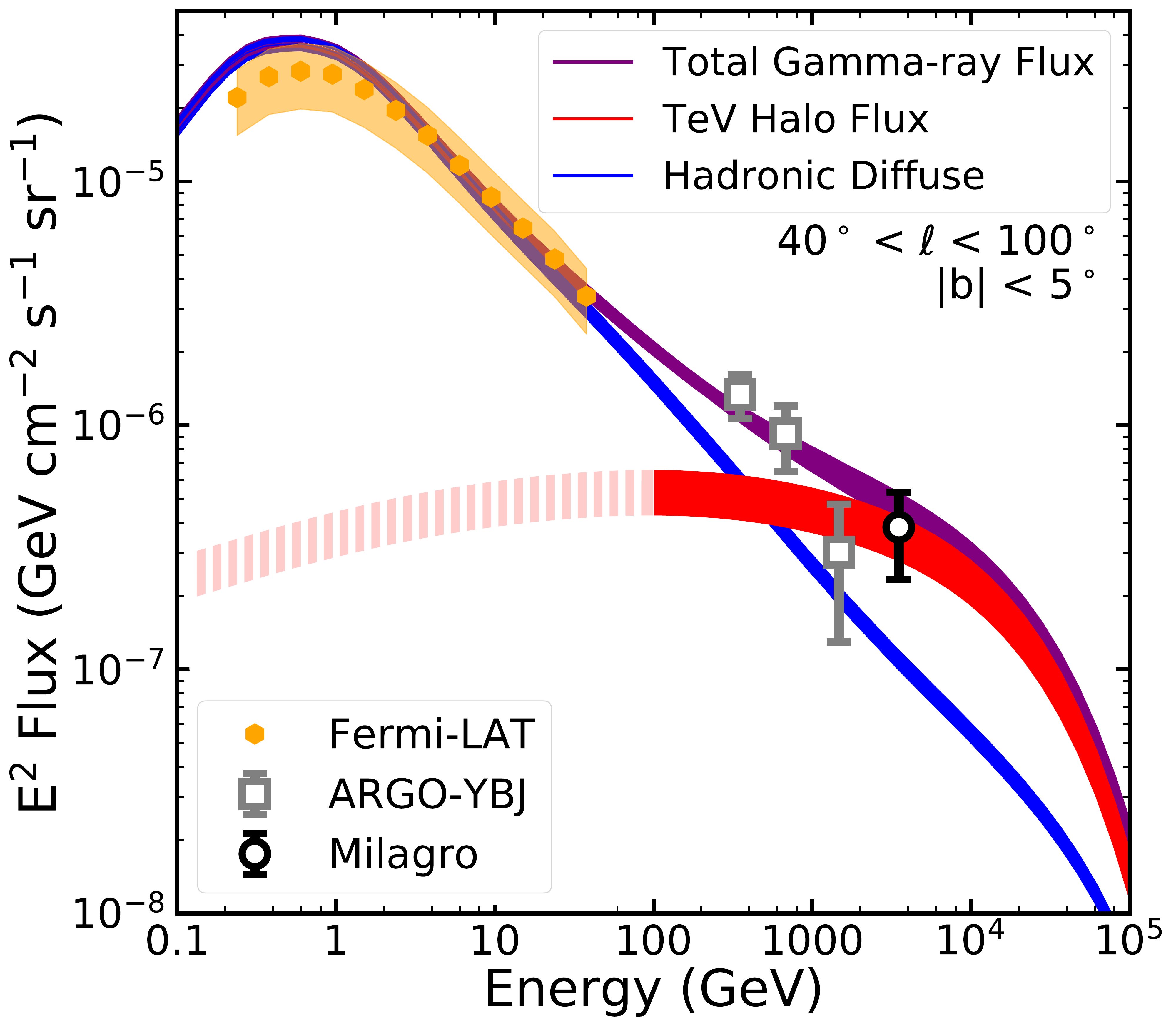}
\caption{The contribution of sub-threshold TeV halos to the diffuse $\gamma$-ray emission along the Galactic plane in the region 40$^\circ$~$<$~$\ell$~$<$~100$^\circ$, and $|b|$~$<$~5$^\circ$, compared to observations by the Fermi-LAT (described in text), ARGO-YBJ \citep{Bartoli:2015era} and Milagro~\citep{Atkins:2005wu}. The background (blue) corresponds to the predictions of 128 {\tt Galprop} models of diffuse $\gamma$-ray emission~\citep{Ackermann:2012pya}, and the diffuse contribution from TeV halos (red) is described in the text. TeV halos naturally reproduce the TeV excess observed by Milagro, while remaining consistent with ARGO-YBJ observations. The dashed red region indicates additional uncertainty due to our ignorance of low-energy e$^+$e$^-$ from TeV halos.}
\label{fig:tevhaloflux}
\end{figure}

We utilize Monte Carlo methods to produce a steady-state pulsar population normalized to the supernova rate and morphology of each {\tt Galprop} model. Each model is, in turn, normalized to the observed distributions of OB stars~\citep{Bronfman:2000tw}, pulsars~\citep{Lorimer:2003qc, Yusifov:2004fr, Lorimer:2006qs}, or supernova remnants~\citep{Case:1998qg}. We calculate the $\gamma$-ray luminosity of each TeV halo following~\citep{Hooper:2017rzt}. Specifically, we pick an initial period at time $t=0$ following a Gaussian with $\mu_p=0.3{\rm s}$ and $\sigma_p=0.15 {\rm s}$, and an initial magnetic field following a Gaussian in log-space with log$_{10}$($\mu_B$/1~G)~=~12.65 and $\sigma_B$~=~0.55~\citep{Bates:2013uma}. We then pick a random pulsar age between 0 and 10~Myr, and spin the pulsar down with a characteristic timescale~\citep{Gaensler:2006ua}:

\begin{equation}
\tau = \frac{3c^3IP_0^2}{4\pi^2B_0^2R^6}
\end{equation}

\noindent where we assume I=10$^{45}$~cm$^2$g and R=15~km. The period of the pulsar evolves following P(t)~=~P$_0$(1+t/$\tau$)$^{1/2}$. The spin-down energy of the pulsar is calculated following \citep{Gaensler:2006ua}:

\begin{equation}
\dot{E} = -\frac{8\pi^4B_0^2R^6}{3c^3P(t)^4}
\end{equation}

We assume that 10\% of the spin-down power is transferred into e$^+$e$^-$ pairs above 1~GeV. This is consistent with our model of Geminga, which indicates that between 7--29\% of its total spin-down power is transferred into e$^+$e$^-$ pairs~\citep{Hooper:2017gtd}. We adopt an e$^+$e$^-$ injection spectrum following a power-law with an exponential cutoff, allowing the parameters $\alpha$ and E$_{cut}$ to vary to fit the Milagro data. We find best-fit values of $\alpha \sim$ 1.7 and E$_{cut} \sim$ 100~TeV. These results are consistent with our model of Geminga, where we found best-fit values of $1.5 < \alpha < 1.9$ and $35~{\rm TeV} < {\rm E}_{cut} < 67~{\rm TeV}$, as well as models of the TeV Halo contribution to the HESS galactic center $\gamma$-rays, where we obtained best fit values of $\alpha$~=~2.2 and E$_{cut}$~=~100~TeV~\citep{Hooper:2017rzt}. We stress that these spectra all provide similar fits to the data, and these results are affected by many uncertainties, e.g: the source-to-source variation in the e$^+$e$^-$ injection spectrum, the cooling efficiency of TeV halos, and the systematic uncertainties in $\gamma$-ray energy reconstruction between different telescopes.

\begin{figure}[tbp]
\centering
\includegraphics[width=.48\textwidth]{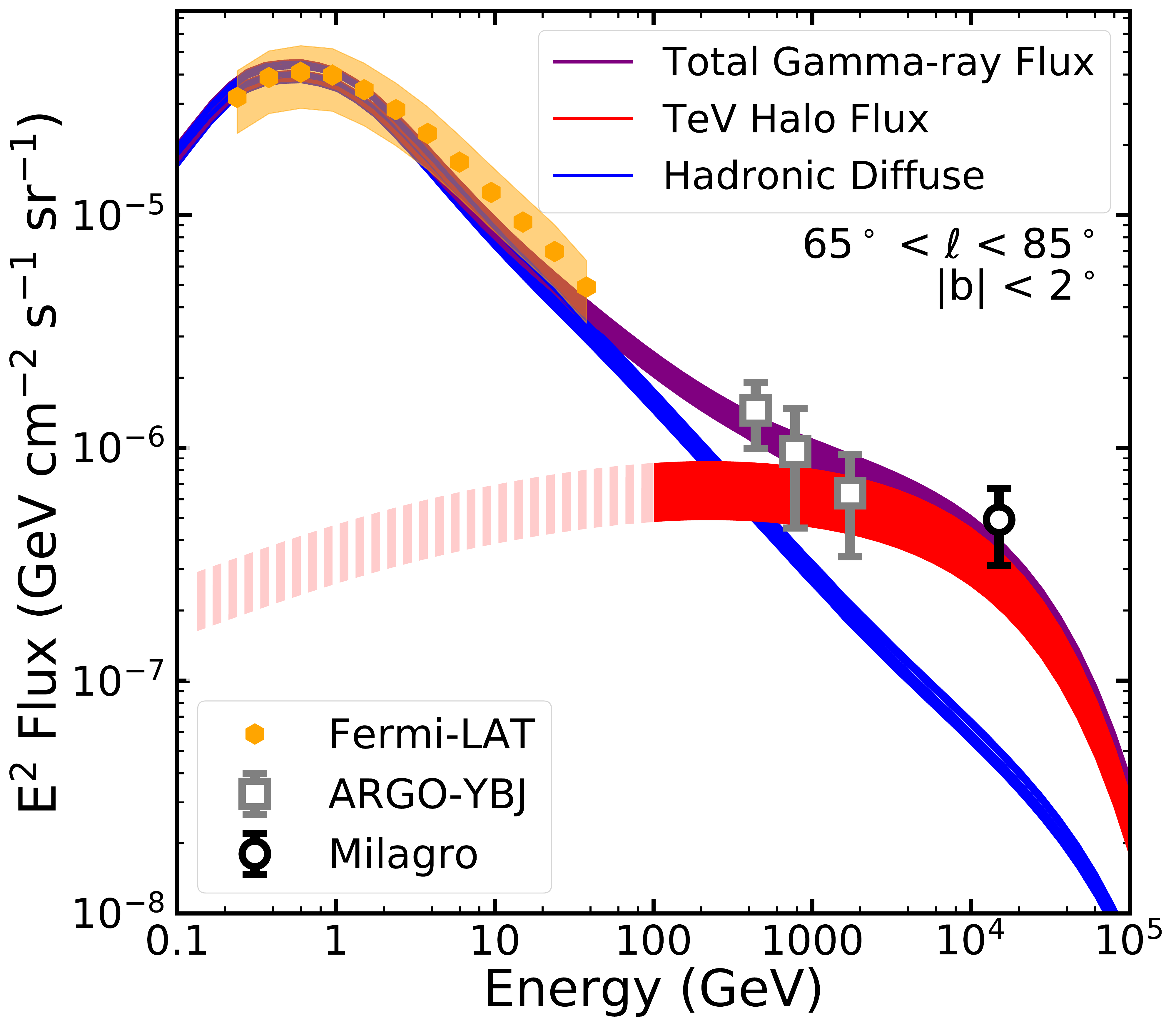}
\caption{Same as Figure~\ref{fig:tevhaloflux} in the smaller region  65$^\circ$~$<$~$\ell$~$<$~85$^\circ$, and $|b|$~$<$~2$^\circ$ examined by Milagro~\citep{Abdo:2008if} at the significant higher energy of 15~TeV, as well as ARGO-YBJ~\citep{Bartoli:2015era}.  ARGO-YBJ observations are quoted in the latitude range $|b|<$5$^\circ$. To correct for this, we renormalize the ARGO-YBJ points based on the ratio of the modeled {\tt Galprop} diffuse emission flux in the $|b|<$2$^\circ$ and $|b|<$5$^\circ$ ROIs. This increases the normalization of the three ARGO-YBJ points by 38\%, 40\% and 42\%, respectively.}
\label{fig:tevhalofluxsmall}
\end{figure}


These electrons are subsequently cooled through a combination of inverse-Compton scattering and synchrotron emission. TeV halos cannot significantly contribute to either the magnetic field energy density or interstellar radiation field (ISRF) energy density over their $\sim$10~pc extent~~\citep{Hooper:2017gtd, Linden:2017vvb}, and thus we adopt values typical of the Milky Way plane: specifically a magnetic field strength of B~=~3~$\mu$G (0.22~eV~cm$^{-3}$), and an interstellar radiation field energy density of 1.56~eV~cm$^{-3}$. We subdivide the ISRF  into a CMB energy density of 0.26~eV~cm$^{-3}$ with a typical photon energy of 2.3$\times$10$^{-4}$~eV, an infrared energy density of 0.6~eV~cm$^{-3}$ with a typical photon energy of 1.73~$\times$~10$^{-3}$~eV, an optical energy density of 0.6~eV~cm$^{-3}$ with a typical photon energy of 0.43~eV, and a UV energy density of 0.1~eV~cm$^{-3}$ with a typical photon energy of 1.73~eV~\citep{Hooper:2017rzt}. 

Unlike the case of individual TeV halos, where e$^+$e$^-$ below $\sim$10~TeV can potentially escape before losing their energy~\citep{Hooper:2017gtd}, the diffuse e$^+$e$^-$ population from an ensemble of TeV halos is further cooled while propagating through the interstellar medium. Assuming a standard diffusion constant of D$_0$~=~5$\times$10$^{28}$~cm$^{2}$~s$^{-1}$ at 1~GV and a Kolmogorov diffusion index $\delta$~=~0.33, e$^+$e$^-$ travel only 0.38~kpc~(E$^{-0.33}$/1~GeV) before losing energy, implying that e$^+$e$^-$ with initial energy above $\sim$50~GeV fully cool before leaving the galactic plane. Thus, we assume that the e$^+$e$^-$ population is in steady state and fully cooled. This assumption fails only at very low energies where the TeV halo contribution is highly subdominant to the hadronic background. Using this cooled electron spectrum, we calculate the inverse-Compton scattering $\gamma$-ray spectrum and intensity, taking into account the Klein-Nishina suppression of inverse-Compton scattering~\citep{Blumenthal:1970gc, Hooper:2017rzt}.

Finally, we note that our Monte Carlo model could (theoretically) produce a single very bright, or very nearby, TeV halo that would dominate the total emission. However, this scenario is ruled out by Milagro, which would have resolved such a source. Milagro barely resolved the extended TeV emission from Geminga~\citep{2009ApJ...700L.127A}, thus we conservatively exclude contributions from any individual TeV halo with a $\gamma$-ray flux exceeding Geminga, (4.27$\times$10$^{-9}$~erg~cm$^{-2}$~s$^{-1}$). We note that our model indicates that only $\sim$1~source brighter than Geminga should exist in our region of interest, and thus eliminating this source is consistent with Poisson fluctuations.

\begin{figure}[tbp]
\centering
\includegraphics[width=.48\textwidth]{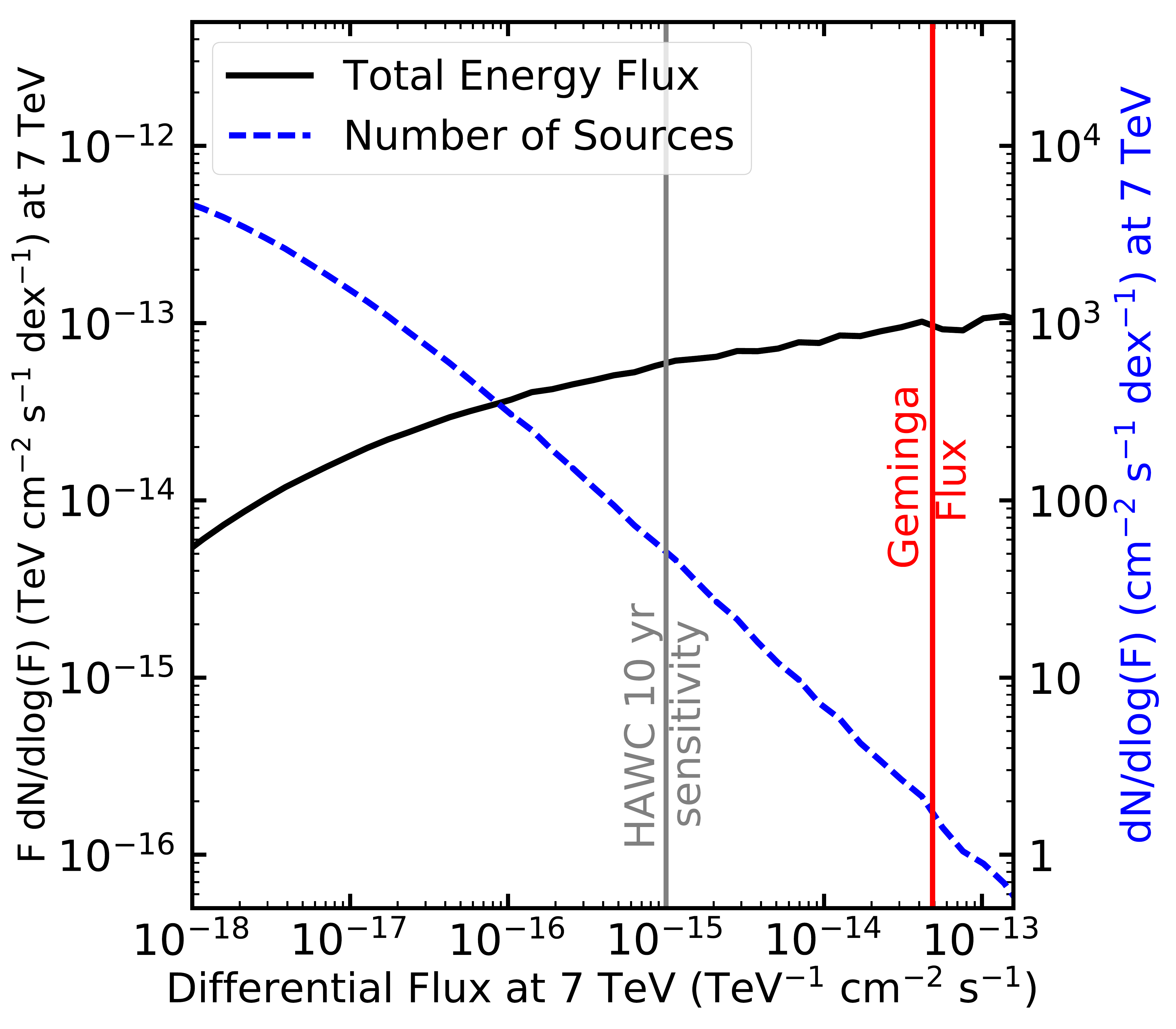}
\caption{Contributions to the TeV excess as a function of the flux of individual TeV halos, modeled in the region 40$^\circ$~$<$~$\ell$~$<$~100$^\circ$, and $|b|$~$<$~5$^\circ$. We normalize our results to 7 TeV~\citep{Abeysekara:2017hyn}, and assume that individual TeV halos convert their spin-down luminosity into 7~TeV $\gamma$-rays with an identical efficiency as Geminga. Vertical lines correspond to the TeV halo flux of Geminga, and the projected 10~yr HAWC sensitivity.  Results are shown for the total $\gamma$-ray flux  (F dN/dlog$_{10}$(F), black, left y-axis), which indicates that a reasonable fraction of the total $\gamma$-ray intensity stems from the brightest TeV halos, as well as for the source count (dN/dlog$_{10}$(F), blue, right y-axis), which indicates that 10~yr HAWC observations are expected to observe $\sim$50 TeV halos in the ROI. For illustrative purposes, in this plot we show the contribution from TeV halos with individual fluxes exceeding Geminga, predicting the existence of only $\sim$1 such system.}
\label{fig:tevhalodistribution}
\end{figure}

\emph {The Gamma-Ray Emission from the Galactic Plane ---} In Figures~\ref{fig:tevhaloflux}~and~\ref{fig:tevhalofluxsmall} we show the key result of this paper. At energies exceeding $\sim$500~GeV, the diffuse $\gamma$-ray flux from leptonic TeV halos becomes significantly brighter than the diffuse $\gamma$-ray flux from hadronic processes. In particular, at the energies of 3.5~TeV and 15~TeV (probed by Milagro) the flux from TeV halos exceeds the standard $\gamma$-ray background by factors of $\sim$3 and $\sim$8, respectively. Additionally, the hard spectrum of TeV halos simultaneously fits both the bright Milagro TeV emission and with the dimmer $\sim$400-1700~GeV diffuse $\gamma$-ray flux observed by ARGO-YBJ. This is intriguing, as an unphysical break in the TeV proton spectrum would be necessary to produce any such feature with hadronic $\gamma$-ray emission. For clarity, we do not show relevant (but less sensitive) results from Whipple~\citep{LeBohec:2000dk}, HEGRA~\citep{Aharonian:2001ft}, TIBET-II or TIBET-III~\citep{Amenomori:2005ud}, noting that our model is consistent with the upper-limits of each study. PeV $\gamma$-ray constraints from CASA-MIA~\citep{Borione:1997fy} and KASCADE~\citep{Schatz:2003aw} may be relevant if we did not include an exponentially suppressed e$^+$e$^-$ spectrum above 100~TeV. We note that this cutoff is both physically motivated by PWN acceleration models~\citep{Gaensler:2006ua} and preferred by our fit to the Geminga data~\citep{Hooper:2017gtd}.

At GeV energies, our model fits the diffuse $\gamma$-ray flux observed by the Fermi-LAT. This is expected, as hadronic emission dominates the diffuse GeV flux, and we use {\tt Galprop} models that have been tuned to Fermi data. To calculate the diffuse GeV $\gamma$-ray flux, we analyze 8.5~yr of Fermi data using standard analysis cuts. We calculate the flux from the Pass 8 diffuse emission model in a binned analysis of the region 40$^\circ$~$<$~$\ell$~$<$~100$^\circ$, and $|b|$~$<$~5$^\circ$, allowing the normalization of all 3FGL sources and the intensity of the diffuse and isotropic components to vary in 0.1$^\circ$ angular bins and five energy bins per decade spanning the range \mbox{189~MeV - 47.5~GeV}. The statistical errors on the diffuse flux are small, and we instead show a 30\% systematic error band corresponding to uncertainties in the effective area and energy reconstruction of the Fermi-LAT~\citep{Ackermann:2012pya}. In the smaller ROI, we find that the limited latitude range makes this analysis difficult, and we instead re-normalize the results from our larger ROI based on the relative diffuse emission intensity at 1~GeV in both models.


To fit the $\gamma$-ray data, we utilized a power-law electron injection spectrum $\alpha$~=~1.7 with E$_{cut}$~=~100~TeV, which is slightly harder than that required to fit HAWC observations of Geminga~\citep{Hooper:2017gtd} or the diffuse galactic center $\gamma$-ray emission observed by HESS~\citep{Hooper:2017rzt}. The necessity for a harder e$^+$e$^-$ injection spectrum is entirely driven by Milagro observation of bright diffuse emission at 15~TeV in the smaller ROI, which is hard to fit with an e$^+$e$^-$ spectrum that is exponentially suppressed at $\sim$50~TeV. We note, however, that the e$^+$e$^-$ injection spectrum in our model is degenerate with both the efficiency of electron cooling in the Milky Way plane, as well as the strength of the interstellar magnetic field, which provides the only alternative cooling pathway for high-energy electrons. Additional observations of TeV halos with well-known spectra will be necessary to determine the range of reasonable values for the TeV halo electron injection spectrum.


In addition to calculating the total diffuse intensity, our model can also determine the individual fluxes of TeV halos that contribute to the Milagro excess. In Figure~\ref{fig:tevhalodistribution} we show the differential contribution to both the total TeV halo number density and the total TeV halo flux as a function of the individual $\gamma$-ray flux of individual TeV halos. We show results for the larger 40$^\circ$~$<$~$\ell$~$<$~100$^\circ$, $|b|$~$<$~5$^\circ$ ROI. Because we care about the emission from individual objects, in this case we do not fully cool the e$^+$e$^-$ spectrum, but instead show the differential flux at an energy of 7~TeV, corresponding to the quoted energy scale of TeV sources listed in the 2HWC catalog~\citep{Abeysekara:2017hyn}. The fluxes of each individual TeV halo are calculated assuming that each source converts the same fraction of its total spin-down power into 7~TeV $\gamma$-ray emission, utilizing the observed flux of Geminga as a template~\citep{Hooper:2017gtd, Linden:2017vvb}. 

We note three critical results. First, our model correctly predicts that $\mathcal{O}$(1) TeV halo as bright as Geminga should exist in the Milagro ROI. In fact, three TeV sources brighter than Geminga are observed by HAWC in this region: 2HWC J2031+415, 2HWC J2019+367, and 2HWC J1908+063. All three sources are best fit by spatially extended $\gamma$-ray templates and overlap the positions of known ATNF pulsars. These observations indicate that they are TeV halo candidates~\citep{Linden:2017vvb}. We note that the latter two sources are coincident with very young pulsars where TeV $\gamma$-ray emission may also be produced by the supernova remnant. Second, we find that 10~yr HAWC observations are likely to definitively prove this correlation, finding $\sim$50~individual TeV halos in the Milagro ROI. Third, we find that a significant fraction of the total diffuse TeV halo intensity is produced by systems that individually exceed 1\% of the Geminga flux. \emph{Our model thus provides a clear and testable hypothesis: a significant fraction of the TeV excess will be resolved into individual TeV halos by future HAWC observations.}

\vspace{0.2cm}
\emph {Conclusions---} 
In this paper, we have assumed that the TeV emission observed from Geminga is typical of young and middle-aged pulsars in the Milky Way. This hypothesis is supported by the observation of $\mathcal{O}$(10) TeV halos with characteristics similar to Geminga, and the lack of any observations which rule out TeV halos in systems where they are expected. In particular, we have assumed that all pulsars younger than 10~Myr convert $\sim$10\% of their spin-down power to relativistic e$^+$e$^-$ pairs, which subsequently cool via inverse-Compton scattering of the ambient interstellar radiation field. These pulsars naturally produce a population of individually sub-threshold TeV halos that power a bright diffuse TeV $\gamma$-ray flux. Intriguingly, the total contribution of TeV halos to the diffuse $\gamma$-ray flux exceeds the total contribution of hadronic cosmic rays from supernovae at energies exceeding $\sim$500~GeV. Moreover, the intensity and spectrum of this emission closely matches the observed TeV excess found in the Milagro data, removing the tension between the soft proton spectrum measured by local cosmic-ray experiments and the hard proton spectrum required by TeV $\gamma$-ray observations~\citep{Gaggero:2015xza}. Since extragalactic $\gamma$-rays are highly attenuated at TeV energies, our results imply that the TeV $\gamma$-ray sky is expected to be dominated by leptonic processes. 

This result has important implications for the source of the very-high-energy neutrinos observed by IceCube~\citep{Aartsen:2013jdh, Aartsen:2015knd}. Rapidly star-forming galaxies (SFG) are a leading source candidate for the IceCube neutrinos~\citep{Loeb:2006tw, Stecker:2006vz, Murase:2013rfa, Anchordoqui:2014yva, Chang:2014sua, Tamborra:2014xia, Senno:2015tra, Emig:2015dma, Bechtol:2015uqb, Moharana:2016mkl, Chakraborty:2016mvc, Linden:2016fdd}. Since no very-high energy emission has been observed from SFGs, their very-high energy neutrino flux must be extrapolated from GeV $\gamma$-ray observations, an extrapolation which assumes a purely hadronic model. Our model indicates that a significant fraction of the TeV $\gamma$-ray flux in SFGs is produced by leptonic TeV halos, which do not produce neutrinos. Thus TeV halos necessarily decrease the predicted SFG neutrino flux. However, the TeV halo contribution at GeV energies is unknown, and multiple uncertainties exist, including: the low-energy spectrum and intensity of TeV halos, the ``calorimetry" of cosmic-ray protons in rapidly star-forming galaxies, and the energy-scale scale that statistically dominates the $\gamma$-ray spectral determination from SFGs. However, we note that current models have found a best-fit SFG $\gamma$-ray spectral index of $\alpha$~=~-2.3, which already implies that SFGs contribute less than $\sim$10\% of the PeV neutrino flux~\citep{Bechtol:2015uqb, Linden:2016fdd}. Even small $\gamma$-ray contributions from TeV halos will continue to make SFG interpretations of the IceCube neutrinos more untenable.  

Finally, we note that this model is imminently testable. In particular, our analysis indicates that most TeV $\gamma$-ray sources are TeV halos~\citep{Linden:2017vvb} --- and we predict that 10~yr HAWC observations will observe $\mathcal{O}$(50)~TeV halos coincident with radio pulsars~\citep{Linden:2017vvb}. Furthermore, these observations will resolve a significant fraction of the TeV excess into individual TeV halos, clearly confirming, or ruling out, the TeV halo origin of the Milagro excess.


\section*{Acknowledgements}
We thank John Beacom, Ilias Cholis, Daniele Gaggero, and Dan Hooper for a number of useful comments which greatly improved the quality of this manuscript. TL acknowledges support from NSF Grant PHY-1404311 to John Beacom.

\bibliography{tevhalos}

\end{document}